**Matters Arising from S. Vaitiekenas et al., "Zero-bias peaks at zero magnetic field in ferromagnetic hybrid nanowires" Nature Physics 2021**

C. Riggert[1], M. Gupta[1], Y. Jiang[2], V. S. Pribiag[1], V. Mourik[3], S. M. Frolov[2*]

[1] *School of Physics and Astronomy, University of Minnesota Twin Cities, Minneapolis, Minnesota 55455, USA*
[2] *Department of Physics and Astronomy, University of Pittsburgh, Pittsburgh, PA 15206, USA*
[3] *JARA-FIT Institute for Quantum Information, Peter Grünberg Institute (PGI-11), Forschungszentrum Jülich GmbH, Campus Boulevard 79, 52074 Aachen, Germany*

\* [frolovsm@pitt.edu](mailto:frolovsm@pitt.edu)

January 8, 2025

In 2021 Nature Physics published a paper by Vaitiekenas, Liu, Krogstrup and Marcus titled "Zero-bias peaks at zero magnetic field in ferromagnetic hybrid nanowires" [1]. The paper reports low temperature transport measurements on semiconductor InAs nanowires with two partly overlapping shells - a shell of EuS, a magnetic insulator, and a shell of Al, a metal that becomes superconducting at temperatures below 1.2K. The paper claims that (1) the data are consistent with induced topological superconductivity and Majorana zero modes (MZMs), and (2) that this is facilitated by the breaking of the time reversal symmetry through a direct magnetic interaction with the EuS shell.

In this Matters Arising, we present an alternative explanation which is based on trivial effects that are likely to appear in the reported geometry. Specifically, first, we find that data the authors present in support of the topological superconductivity claim can originate from unintended quantum dots in their devices, a widely known likely explanation that is not being discussed in the paper. Second, our analysis of the setup, supported by our numerical micromagnetic simulations, shows similar effects could be obtained due to stray magnetic fields from the region of the EuS shell damaged during Al etching. As we show, the presence of stray magnetic fields is highly likely in this geometry under the processing steps employed. These stray fields suppress superconductivity in the Al shell locally, and induce Zeeman splitting in an unintended quantum dot formed where the nanowire was etched, causing trivial subgap states in the quantum dot to move closer to zero bias voltage. This would account for zero-bias peaks at zero field shown in the paper, and does not require topological superconductivity. This basic picture should come before the exotic interpretation in terms of magnetic exchange interaction with a ferromagnetic insulator [2].

The authors discuss their findings as "topological superconductivity" and "Majorana zero modes (MZMs)" throughout the paper. However, since 2014, when Lee et al. studied quantum dots coupled to superconductors in the presence of magnetic fields[3], the nanowire community has been well aware of the ability of trivial states to mimic MZMs. In fact, this interpretation of the data, sometimes referred to as 'alternative' with respect to the MZM interpretation, is likely behind most if not all reported zero-bias peaks in hybrid nanowire devices and beyond[4–6]. The authors of Vaitiekenas et al. did not include the 2014 paper or any subsequent experimental reports as references, to the effect that a reader of their paper may come away unaware of this.

More specifically, Lee et al. demonstrate that a quantum dot coupled to a superconductor can exhibit near zero-bias tunneling conductance peaks when the gap in the lead is collapsing. This is because the states above the gap push the trivial quantum dot subgap states to lower energies. The manifestation of these trivial states considered by Lee et al. includes a dot with tunable coupling to the superconducting lead, ranging from the weak to strong-coupling regimes. Strongly coupled dots often do not display well-defined Coulomb diamonds and the subgap states are located closer to the gap edge at zero magnetic field, but move closer to zero bias at finite fields, which would include stray fields from a nearby micromagnet.

In a follow-up on Vaitiekenas et al, similar data were obtained by some of us, including hysteretic supercurrents and zero-bias peaks in conductance near zero applied magnetic field [7]. However, in that experiment, by its design, no MZM were present. Furthermore, magnetic materials were not in contact with the nanowire. Instead, stray magnetic fields were coupled to trivial states in gate-controlled quantum dots. The great interest generated by the Vaitiekenas et al 2021 publication [8–14], on the background of the lack of discussion of trivial explanations, along with our own data that disfavor topological superconductivity based on such evidence, motivated our careful re-analysis of their experiments.

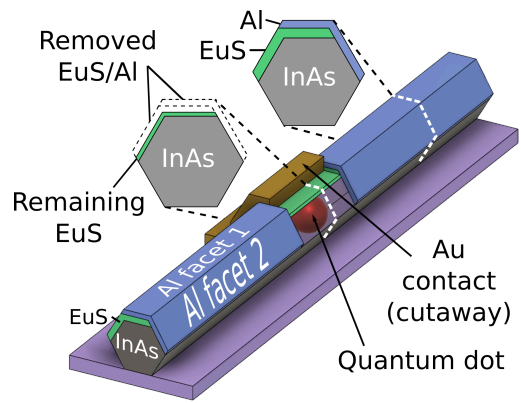

**Figure M1:** *Schematic of the device showing the unintended quantum dot in the regions of the etched shell.*

Evidence of trivial quantum dot states

We find that an unintended quantum dot located within the authors' nanowire can explain the data in the figures of the Vaitekienas et al. paper, without the need for exotic interpretations. Figure M1 provides a schematic previewing our explanation.

In Vaitiekenas et al., Nature Physics 2021[1], the authors presented in Fig. 1 of that paper supercurrents through the Al shell on an InAs/EuS/Al nanowire. These supercurrents are hysteretic in the external magnetic field and the authors argue this confirms that magnetism in

EuS influences superconductivity. That device underwent nanofabrication to add top contacts. The paper then presented in Figs. 2-4 tunneling spectroscopy including zero-bias conductance peaks from a different nanowire device, where in addition to contacts a junction was etched in the middle of the nanowire.

Fig. 2a of Vaitiekenas et al., Nature Physics 2021 contains an image of device 2, in which the authors had etched the shell of the InAs nanowire at the position of the gate, labeled "$V_C$". Our inspection of this image revealed that the shell was not uniform along the nanowire which can be seen from the wiggles in the blue-colored region; the boundary of the Al shell is forming an irregular shape. There is not enough data to understand what has caused the inhomogeneity, however Al layers do tend to coagulate and de-wet upon heating during the lithography cycles. In addition to the shell itself being inhomogeneous, in devices with etched shells quantum dots form near the etched region due to inhomogeneous potentials. Unintentional quantum dot effects were observed in the overwhelming majority of devices with related geometries. [15]

The authors' experimental data in Figs. 2, 3, 4, and their Extended Data Figs. 4-10 are fully consistent with subgap states originating from quantum dots. The data do not contain any evidence pointing exclusively or preferentially towards the MZM origin of the resonances, nor favoring the MZM interpretation over the more commonly occurring quantum dot interpretation. The evidence of quantum dots is visible in the data as presented. For instance, in Fig. 2b, the bright line below which the authors marked a dash labeled "d" and the broader lines parallel to it above the dash, indicate quantum dot states tunable by both gates $V_C$ and $V_{BG}$.

Near-zero-bias peaks from device 2 presented in Figs. 2c and 2d are largely localized in a small region of Fig. 2b, defined by horizontal and vertical dashes, which is within a single charge state of a quantum dot, though device 2 likely had more than one unintended dot near the etched region. Other features in the data confirm the quantum dot origin of the phenomena: throughout the panels in Fig. 2, the numerous brighter lines present above the induced gap in all the panels are attributable to quantum dot states. Some of these created subgap states that were strongly coupled and localized near the edge of the induced gap, producing the observed loop-like structures when coupling to lower-lying subgap states. Still others are difficult to discern due to the color scheme in Fig. 2, but can be made-out crossing the gap.

We further note that superconductivity is severely suppressed because the apparent superconducting gap feature is near 50 microvolt, when compared with the typical Al gap of 200 microvolt. This means that trivial quantum dot states would be pushed to zero bias by the collapsing gap, precisely as seen in Lee et al. in 2014[3]. This would account for the appearance of zero-bias peaks induced by trivial quantum dot states. The characteristic splittings of zero-bias peaks observed at the edges of panels Fig. 2d and 2e are tell-tale signs of trivial subgap states.

Noting the data presentation, we remark that zero-bias peaks presented in Fig. 2d appear stretched, in part due to the choice of panel shape and axis scaling, rather than their actual extent in the range of relevant parameters such as the gate voltage. This can be seen from the quantum dot resonances that criss-cross the image having a shallow slope. The result is the impression that the near-zero resonances are more robust than they are.

Stray fields from a damaged EuS shell

Trivial quantum dot states were previously reported to form zero bias peaks at finite magnetic fields. Here the authors show zero-bias peaks at zero applied field. They attribute these to an exotic interaction with the magnetic insulator shell through direct contact with either Al or InAs. They argue that their earlier measurements on undamaged shells did not reveal magnetic domains that would cause stray fields from the EuS micromagnet to spill out into the nanowire[16]. In a single-domain shell, the stray fields will be concentrated around the ends and will not be present in the middle of the nanowire. However, in a damaged shell the etch or mill zone at the junctions or at the contacts can nucleate stray fields at that position. Inhomogeneous superconductor layers can contribute to the formation of magnetic domains and stray fields as well.

The paper reports devices fabricated by first adding a shell of EuS on 2 of the 6 facets of the nanowire. Then a shell of Al was added on 2 facets, with one of the facets overlapping between the two shells. Finally, a junction was etched, primarily in the Al shell by wet etching, and the contacts were made using lithographic steps which include acids, bases, high temperatures and mechanical damage. Even if the etch was highly selective to Al, some damage likely occurred to the EuS shell, either by the incorporation of Al into EuS due to heating, by incomplete removal of Al creating inhomogeneities at the junction, residues due to wet processing, or by partial removal of EuS. In our micromagnetics simulations, we study the stray fields that appear when reducing the thickness of the EuS shell over the junction. Other ways in which nanofabrication or growth can disrupt the integrity of shells are likely to also lead to stray fields over the junction.

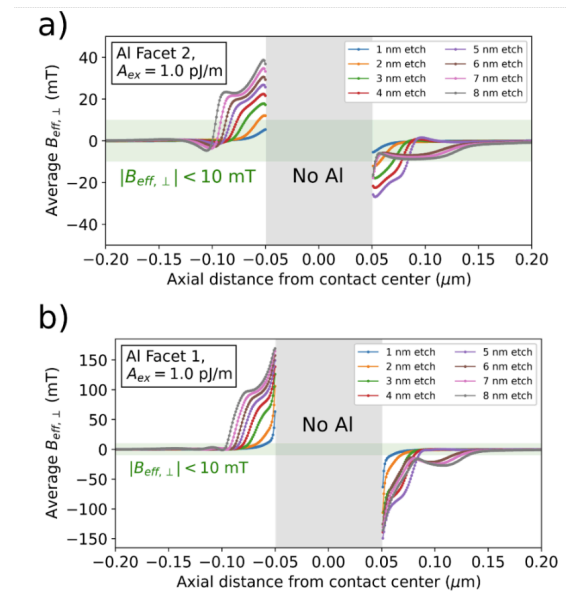

**Figure M2. A and B:** *Facet-averaged normal field to both Al facets for various degrees of etching, showing that any amount of etching is sufficient to generate superconductivity-suppressing normal fields near the contact. Green band shows fields perpendicular to the facet below 10 mT, where we do not expect suppressed superconductivity.*
*Simulation is performed using MuMax3, exchange stiffness $a_{ex}$=1.0 pJ/m and a 1 nm cell size in a 1024 nm long wire with a central 100 nm wide break in the aluminum and partial etch in the EuS.*

We find that, for both Al facets, reduced EuS layer thickness produces stray magnetic fields, locally near the junction, that are large enough to either fully or significantly suppress superconductivity in Al (Figure M2). Suppressed superconductivity would collapse the induced superconducting gap and force the trivial quantum dot states to move towards zero bias eventually forming zero-bias peaks in tunneling spectroscopy.

Typical out-of-plane (perpendicular) critical fields for Al films are in the low 10's of milliTesla. Our results show that comparable local stray magnetic fields appear near the etched junction for the minimum EuS layer etch (by 1 nm out of the 8 nm thick shell). The stray fields are strongest in the Al Facet 1 which is on top of EuS. The fields are also significant in Facet 2. The total suppression of the gap is harder to model because it involves inhomogeneously suppressed superconductivity and the proximity effect. However, it is clear from the micromagnetic simulations that stray magnetic fields due to inhomogeneities in the EuS shell can cause significant suppression of local superconductivity. The region of the strongest suppression of superconductivity is of order 100 nm near the etched junction. This is the region where the spurious quantum dot is apparent in the tunneling spectroscopy in Vaitiekenas et al, Nature Physics 2021 (See supplementary Figure M5 for fields farther away from the junction).

These results show that it is not necessary to consider exotic types of magnetic coupling, such as through direct exchange interaction, in order to account for the reported features in the Vaitekienas et al. paper. Much more trivial explanations can provide similar features and are more likely to arise in practice than the exotic explanations. The findings further support the interpretation of transport data in terms of trivial states localized to quantum dots. Some of the original authors have published an earlier paper arguing that their shells are single-domain, based on scanning SQUID microscopy [16]. But those were as-grown wires which did not contain etched junctions and did not undergo extensive nanofabrication processing. We provide the analysis of domain structure in etched EuS layers using two different exchange stiffness values in the Appendix, with a discussion.

Finally, we note that our analysis is based on the data presented in the paper itself and its supplementary information. If the authors of Vaitiekenas et al., Nature Physics 2021 have more data, especially such that reinforce their favored interpretation, we encourage them to share these data. Likewise, if their unpublished data from these experiments would add evidence to our interpretation of their experiments, we request that they share any such data.

**Acknowledgements.** Numerical simulations for this work were done under the Department of Energy award no. DE-SC-0019274.

# Appendix: Additional micromagnetics simulations

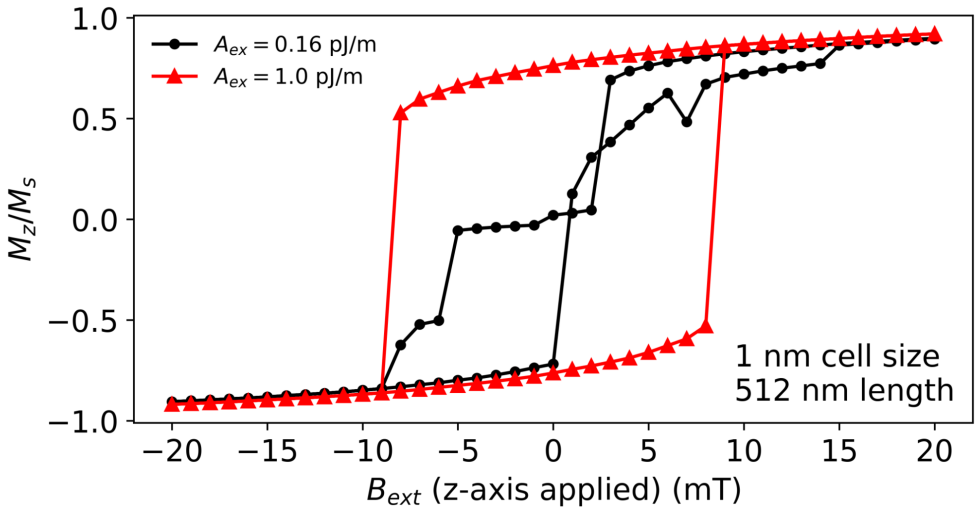

**Figure M3**: *This is a hysteresis loop for a 512 nm length of wire with both the literature value ($a_{ex}$ = 0.16 pJ/m) and our value ($a_{ex}$ = 1.0 pJ/m) of the exchange stiffness, using a 1 nm cell size for both. This demonstrates that an enhanced stiffness is required to obtain clean switching at approximately the values seen in Vaitiekenas et al.*

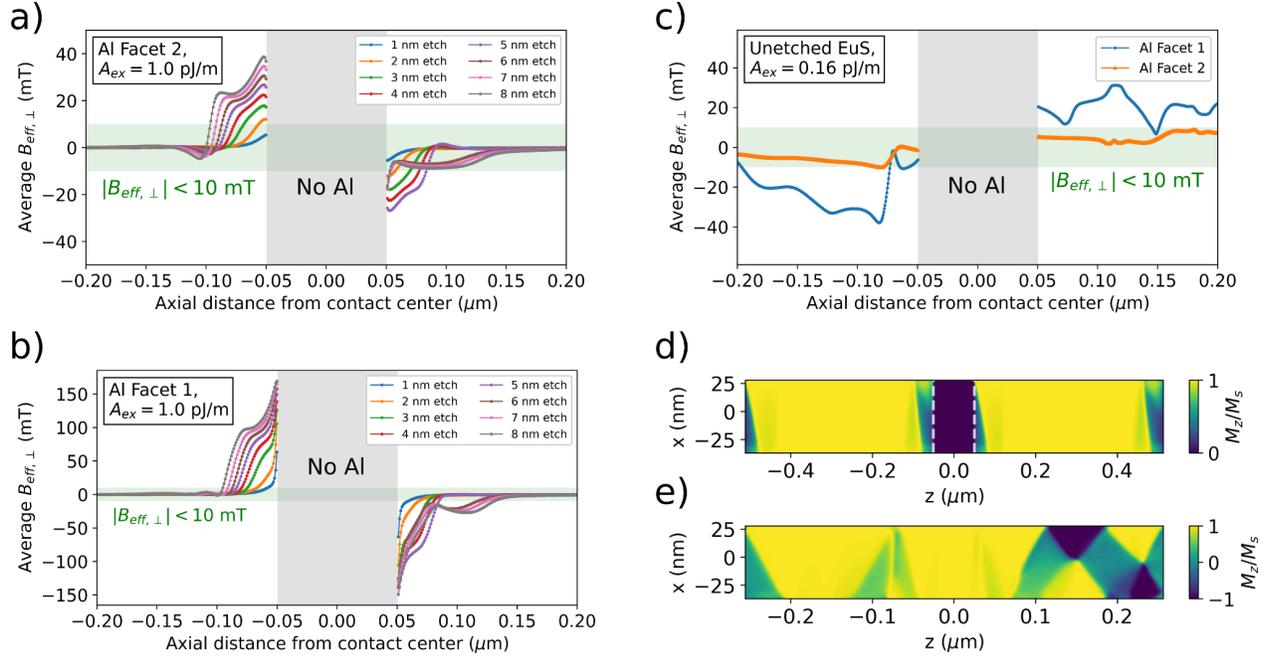

**Figure M4:** *This figure demonstrates that at either level of stiffness, stray fields can be generated which interfere with superconductivity in the Al shell. In panels A and B we show the facet-averaged normal field to both Al facets for various degrees of etching, showing that any amount of etching is sufficient to generate superconductivity-suppressing normal fields near the contact. These are done with $a_{ex}$=1.0 pJ/m and a 1 nm cell size in a 1024 nm long wire with a central 100 nm wide break in the aluminum and partial etch in the EuS. In panel C, we consider an unetched EuS shell with $a_{ex}$=0.16 pJ/m. We show that stray fields from domains in the $a_{ex}$=0.16 pJ/m magnet are sufficient to suppress superconductivity even when there is no damage to the EuS shell. Here a 0.5 nm cell size is used, with the simulated wire length as 512 nm. In panel D, we show a slice down the middle of the top facet of the EuS from panel A and B ($a_{ex}$ = 1.0 pJ/m) to illustrate the formation of clean closure domains and a domain free bulk. In panel E, I show an equivalent slice of the magnet from panel C ($a_{ex}$ = 0.16 pJ/m) to show the presence of domains throughout the bulk, creating the stray fields seen in panel C. All panels are generated by applying a 50 mT axial field, relaxing the magnet, removing that field, and relaxing again, so as to simulate reasonable experimental conditions.*

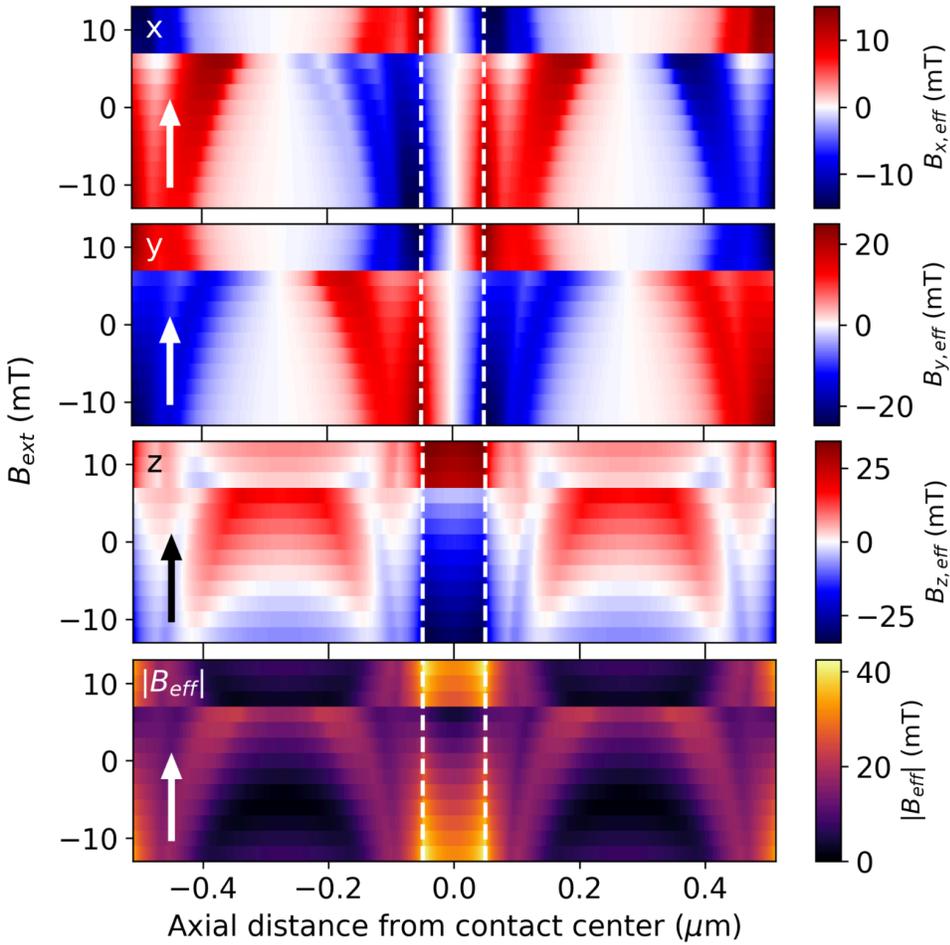

**Figure M5:** *In this figure we consider a 1024 nm length of wire with $a_{ex}$ = 1.0 pJ/m, 1 nm cell, and a full etch of the EuS at the 100 nm wide contact (marked by the white dashed lines. We average the effective field over the wire cross section as a function of length along the wire. We plot this average both component-wise and as a magnitude for a sweep of the external field (along the length of the wire) from negative to positive. This shows both clean switching across all components and the magnitude, as well as a substantial effective field in the portion of the wire beneath the contact, where an incidental open quantum dot is likely to appear and host trivial subgap states.*